# High-concurrency Custom-build Relational Database System's design and SQL parser design based on Turing-complete automata

Wanhong Huang, DLUT
mannho_dut@yahoo.co.jp

## Abstract

*Database system is an indispensable part of software projects. It plays an important role in data organization and storage. Its performance and efficiency are directly related to the performance of software. Nowadays, we have many general relational database systems that can be used in our projects, such as SQL Server, MySQL, Oracle, etc. It is undeniable that in most cases, we can easily use these database systems to complete our projects, but considering the generality, the general database systems often can't play the ultimate speed and fully adapt to our projects. In very few projects, we will need to design a database system that fully adapt to our projects and have a high efficiency and concurrency. Therefore, it is very important to consider a feasible solution of designing a database system (We only consider the relational database system here). Meanwhile, for a database system, SQL interpretation and execution module is necessary. According to the theory of formal language and automata, the realization of this module can be completed by automata. In our experiment, we made the following contributions: 1) We designed a small relational database, and used the database to complete a highly concurrent student course selection system. 2) We design a general automaton module, which can complete the operation from parsing to execution. The using of strategy model and event driven design scheme is used and some improvement on general automata, for example a memory like structure is added to automata to make it better to store context. All these make the automata model can be used in a variety of occasions, not only the parsing and execution of SQL statements. Our complete source code has been uploaded to GitHub.*

## 1. Introduction

In modern software design, almost all projects need to manage and store data. Therefore, the use of database system is essential for almost every project. General database systems, such as MySQL and Oracle, have been widely used in many projects. In order to achieve higher customization and provide storage efficiency for specific projects' data, we need to design a special database in some cases. It is challenging to design a database system. It needs to be applied to a variety of subjects and engineering knowledge. For example, to design a SQL parser, we need the theory of formal language and automata. To improve the efficiency of data access, we need to understand the relevant principles of the operating system, as well as the implementation of cache. How to keep the availability, data consistency and efficiency of the database system under high concurrent development needs careful consideration. Message queuing, distributed caching and other technologies can be used. At the same time, the distributed environment will make the problem more complex.

In our experiment, we will design a small database system and use it in the students' course selection system. And assume that the students' course selection system needs to meet the needs of high concurrency. At the same time, test the data consistency and integrity under the high concurrent.

In order to complete the project, some necessary knowledge is listed below.

### 1.1. Formal Language and Automata

Formal language is different from natural language. It is a language system designed by human beings to complete some work. For example, programming languages, regular expressions, etc. Language is a collection of texts that grammar can recognize. According to the classification, there can be regular languages, context free languages, context sensitive languages, and enumerating recursive languages. RE language includes the other three languages, corresponding to the Turing machine for recognizing the language. Even if SQL Language is not a Turing complete language, we expect to build a Turing complete model to implement it in this experiment.

$$A \rightarrow a, A \rightarrow aB, C \rightarrow \epsilon$$

Figure 1. The grammar production of a Regular Language can only be the above form.

$$A \rightarrow \alpha$$

Figure 2. The grammar production of a Context Free Language can only be the above form. And $\alpha \in (V + U)^*$, V is a set of variables, U is a set of terminals.

$$\alpha A \beta \to \alpha \gamma \beta$$

Figure 3. The grammar production of a Context Sensitive Language can only be the above form. And $\alpha, \beta \in (V + U)^*, A \in V$, V is a set of variables, U is a set of terminals.

For recursively enumerable language, it is a formal language for which there exists a Turing machine (or other computable function) which will enumerate all valid strings of the language.[1]

## 1.2. Cache and Paging

In order to improve the efficiency of data access, we often need caching and paging mechanism. Paging and caching mechanism makes use of the spatial and temporal locality of data, that is, when a certain data is accessed at a certain time, it is likely to need to access it again in a period of time in the future. When accessing the data of a certain location, it is likely to use the data near it. A page is a collection of multiple pieces of data. It will be put to cache as a basic unit. When cache is full, page replacement algorithm is needed for a new page. Commonly used page replacement methods are FIFO, LRU, LFU, etc.

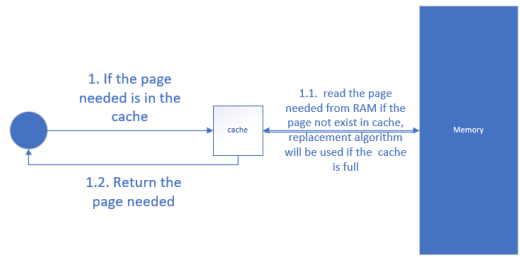

Figure 1. A simple process of caching.

The use of cache technology can greatly speed up our reading speed, because the use of cache makes the interruption of I/O operation greatly reduced. RAM speed is much faster than I/O speed.

## 1.3. Lock

Lock technology is to solve the problem of data consistency under multithreading. For example, when B completes its operation and is about to write back the data, the time slice reaches thread A. Then thread A reads the same data and the data it read is the data that still not been updated, which causing an error. In our experiment, we will use the read-write lock. It can allow multiple tasks to read the same object, but at most one task to write this object, and read and write is mutually exclusive.

For a database system, the objects we want to lock are usually tables or tuples(rows).

In InnoDB, the default lock level is row level lock. But the whole table will be locked if use the statement that *select … for update.* And InnoDB supports shared lock and exclusive lock. Shared locks is lightweight lock. It is suitable for weak competition conditions. It assumes that competition does not exist. When is about to write the data back, if it is found that the current data's version is different from the previously read data's version, the transaction will be rolled back.

## 1.4. Multithread and Thread Pool

In order to ensure the efficiency of data access and meet the requirements of high concurrency environment. Multithread is a must. It can make server capable to process multi tasks in parallel. However, the introduction of multithreading mechanism will also bring some problems, which has been introduced in 1.3. And to improve the efficiency of the system and improve the reliability and availability of the system, Thread Pools are expected be used on the program. Because the thread pool can prepare the threads in advance, so that when a request comes, it can wake up a thread quickly instead of creating a thread. The time cost of creating a thread is much greater than that of waking up a thread.

Furthermore, thread pool can also limit the number of threads. Each thread has its own stack space, and each thread will occupy a certain amount of space. In the case of high concurrency, we sometimes want to limit the number of threads to prevent excessive space consumption.

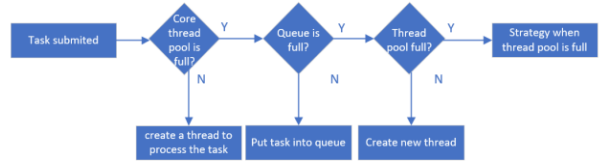

Figure 2. Java thread pool mechanism

There are mainly 4 kinds of thread pool in Java SE. They are *Cached Thread Pool*, *Fixed Thread Pool, Scheduled Thread Pool and Single Thread Pool.* We can also use *ThreadPoolExecutor* to create a more complex thread pool. The details will be discussed lately.

## 1.5. Index and B+ Tree

Index is a way to optimize query speed. A tree structure is used to store the index attribute values, and the corresponding contents can be found in the logarithmic progressive time complexity. There are many kinds of trees can be used to establish index such as ALV, Red Black Tree, B, B+, B-, B*. Usually B+ tree has a good performance on storage of database index. Because in B+ tree, the real object pointer only exist in the leaves which can save a lot of space. Furthermore, All leaf nodes form an ordered list. When you reach a leaf node, you can access other node orderly.

### 1.6. Socket API

Socket is the abstraction of two-way communication between application processes on different hosts in the network. OS or some libraries has provide with a set of socket API for us to make communication between hosts easily. Socket API is also IO operation in essence. In order to improve efficiency, we should also optimize it. There are three I/O types for us to choose: BIO(Blocking I/O), NIO(No-Blocking IO) and AIO(Async IO) .

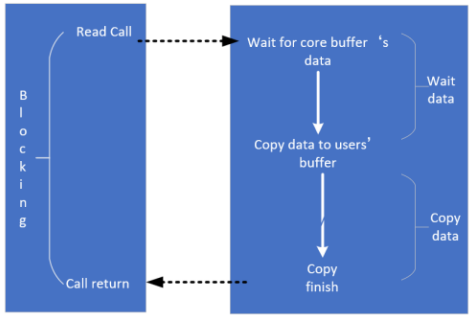

Figure 4. BIO's process. In this way, the caller thread must block itself though no data in the buffer. and wait until the data transfer is completed. The thread can not do other things during the data transfer process.

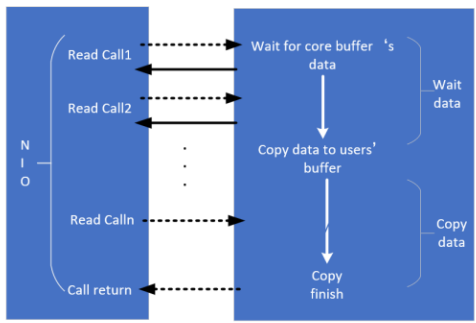

Figure 5. NIO's process. In this way, the caller thread will try to read the data buffer many times, it will return to do other task if there are not data in the buffer, and call again to see if can read some data from the buffer.

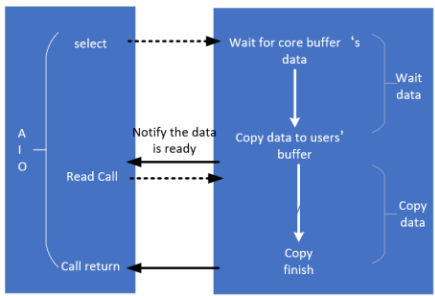

Figure 6. AIO's process. AIO shows a more efficient performance in high concurrence environment. The caller thread not frequently judge if the buffer has data, but let selector to notify it the data is ready and can begin read operation.

### 1.7. Proxy and Database's log

By using proxy design pattern, we can easily realize the log function, because a proxy interface can plus some work before or after the target functions. And We need logs to reduce the impact of accident.

### 2. Requirement Analysis

First of all, as a relational database system, it should be able to achieve the core functions including table management, data's addition, deletion, modification and query. All operations should be done through SQL statements. Secondly, we must ensure the access efficiency and data integrity in high concurrency environment. Integrity constraints are required. It should able to create indexes to optimize query speed.

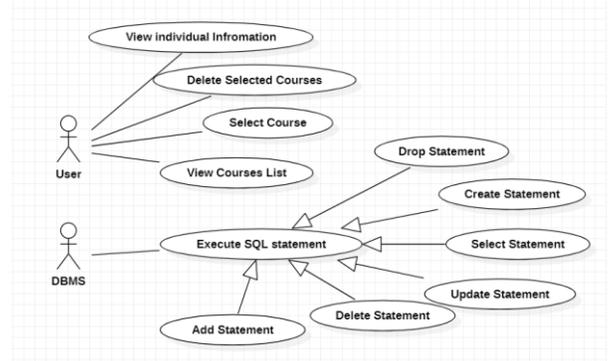

Figure 3. UML Use Case Graph

In a word, the design of our DBMS should meet the speed requirements as much as possible, and be able to deal with high concurrency tasks, as well as save availability and data integrity in high concurrency environment. Besides. The database system should be able to access directly through socket API, so that the database system can be accessed by using any programming language. For security, we can design our own application layer protocol by using the socket API.

### 3. Overall Design

We have two programs in this project. One is DBMS, which will receive the request from network and analyze the SQL statement in the request, and then process the target operation for the client, finally return a result to the client. It's the core part in this project. Because it is responsible for the SQL statement analysis, Cache and indexes' management and all the operation to the file system

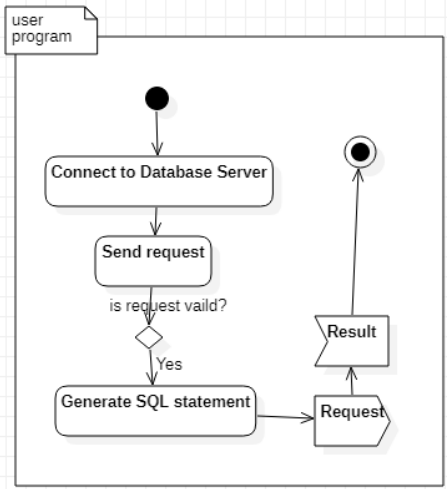

Figure 7. A simple UML Activity Diagram of user program.

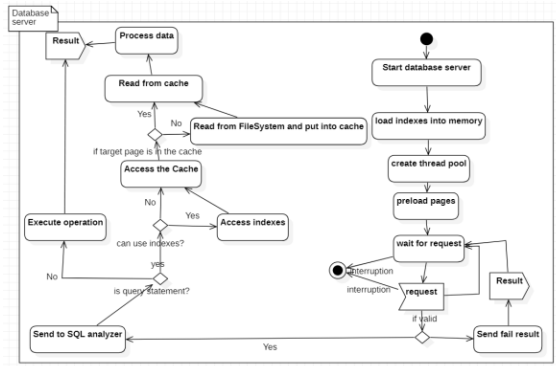

Figure 8. A simple UML Activity Diagram of database server.

The simplified activity diagram of user program and database server program is located in Figure 7 and 8. It should be noted that in this project. We chose to load all indexes into memory when the database server program started. Because in this project, the DBMS we designed is more concerned with speed. Whether or not to use cache storage for db's indexes depends on the specific business. This is also the advantage of self-made database system. We can speed up specific business in many details.

4. Implementation

4.1. Design of Turing-Complete Automata

To analyze and execute SQL statements, we need at first make a SQL parser, which can usually be made according relative Automata and Formal Language theory. In fact, general SQL language is not a Turing complete language.

We can simply using PDA(Push Down Automata) to make a grammar analyzer. But in this project, we choose to make some change in common automata, can create a general automata, which can finish all the task in SQL statements' execution.

Same to the general automata, We first need create a Automata Node class to represent Automata's Node. Then we create a Automata class. Each instance of it represent an entire automata. Different from PDA, which maintain a stack to stock context information and it can analyze Context Free Language but can't tackle with Recursively Enumerable Language. We put a hash table to the context, which likes a storage. If you put a stack object into it, the automata will become a PDA. As stack restricts the output order of data. By using a storage like structure, it can become more flexible. We can put different kind object in to the storage to satisfy different task.

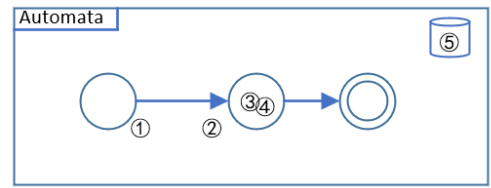

Figure 9. A simple event driven automata graph. We added a key-value storage structure (④) to the automata object. And when the state is to be shifted through an edge, the corresponding events will be triggered before(①) and after(②) the shift. You can handle the context of the automaton, or even the automaton itself.

The automata can becomes Turing completing because this automata is fully event driven and the node function delegation interface is put into AutomataNode class. State Changing and State Changed Event interface is also declared in Automata Edge. We reach a node, it we also trigger a Node Reached Event, in which has a default realization that call the node function delegation to do some work. You can also rewrite it to make it call the node function delegation conditionally. And you can easily write a endless loop code, which usually means it is a Turing-complete structure and it may never halt. You can also use this structure to try out many things. Even changing the automata structure in the process of state transition. Because the programming language we use is Turing complete. In theory, anything can be done.

By collect the necessary information in the grammar parse process, we can execute the final operation in a specific node.

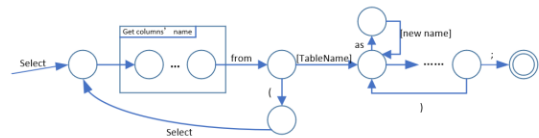

Figure 10. An example of recognize *select statement.*

The automata in figure[10] is a simple automata that can recognize select statement (For simplicity, some parts are omitted). Suppose we want to parse a statement that *select c1,c2 from (select \* from TA) as tmp;* It's a select statement that include sub-select statement.

The parsing process is shown in figure [11][12][13].

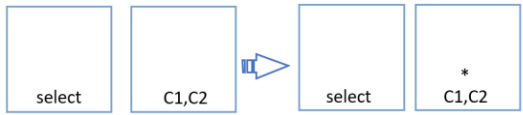

Figure 11. The contents of storage during parsing process.
Left: after reaching the first left bracket
Right: after reaching the * in sub-select statement

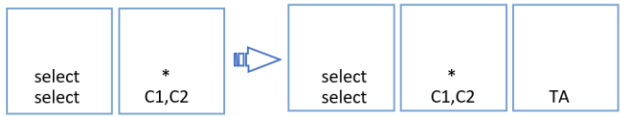

Figure 12. The contents of storage during parsing process.
Left: after reaching the second *select* keyword
Right: after reaching the table's name TA

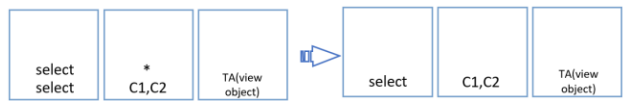

Figure 13. The contents of storage during parsing process.
Left: after the end of subquery, from the table name TA it will create a view object, which includes data of TA.
Right: after the end of subquery, remove the information of subquery and become 1-level *select* statement now.

As every node function will return an Object type object. The query result will be wrapped in a view object and return from the terminal node.

This automata model is theorical Turing completed. For example you can construct a automata by this model to execute a dead cycle operation.( Figure 14. )

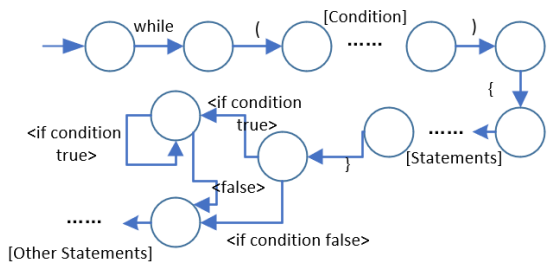

Figure 14. A automata can parse *while* statement and realize dead cycle. When receive *while,* it push a while flag to a stack in storage. Then after receive condition expression, it will create a Condition object and put it into the storage. After receive } , in Edge_BeforeShift event of next two edges, it will check the condition object and judge if the state can shift through this edge.

### 4.2. Physical Layer

For the sake of simplicity, the data is directly stored on the ordinary text file without coding, and the data is not compressed. (otherwise, decompression algorithm will takes time when reading. We expect this database can be faster.)

For the string type, we used fixed length string. The maximum string length need to be defined when creating a table. Besides, the tables structure and metadata will be storage independently in a table structure file. As for the index. It is also storage in a separate file. The index's key is index column data with values of the corresponding record number. The file organization is shown in Figure 15.

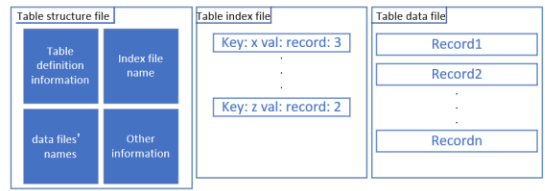

Figure 15. Files organization

Before database server starting, structure of all the tables will be load to memory as a LogicTable object. For the read/write of a table, it should completed through a TableReadWrite class (A singleton class) which maintain TableReader object of each table and TableWrite object of each table. ReadWriteLock is used to prevent inconsistence in high concurrent occasion.

### 4.3. Cache Management

For faster reading the data, we are using cache in this program. (For efficiency, we only used cache for table data. Ass for index, it will be fully put into memory in advance. For large data, we should also create cache for index data. However, in our project, our database has higher requirements for speed, assuming that the memory space is sufficient. So choose to fully load index to memory.).

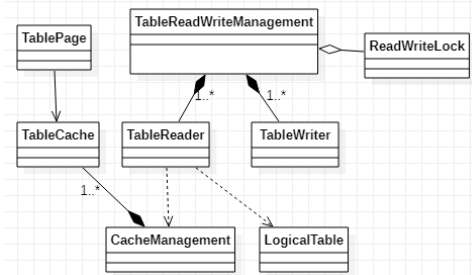

Figure 16. A simple UML class diagram of the several class mentioned.

To manage the cache, a CacheManagement class is needed, which is also a singleton class. Every table has its own cache. CacheManagement class maintain serval cache of different tables by a hash map.

Several records will be storage on page object, and the page size is defined when creating a page object. The page will be stored in the cache as the base unit. We can find the corresponding record by record number. And find the page number through record number and page size.

### 4.4. Index Management

Using index is an important way to improve our efficiency of reading data. We choose B-plus tree as the structure to storage our index. So we can directly using the hashmap object in Java, which is realized by Red-Black tree. We wrote a B-plus tree class in our project. A singleton class IndexManagement is also made to maintain index of tables.

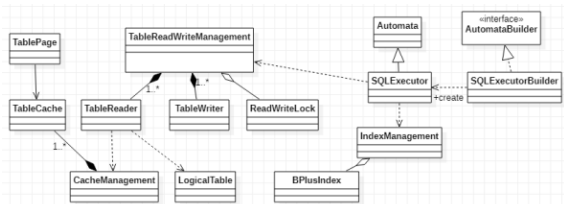

Figure 17. A simple UML class diagram of the several class mentioned.

When SQL executor meet a *select* statement, it will find if it can use index to promote query speed. And try to access cache after getting a record number from index.

### 4.5. Integrity Constrain

Integrity constraint is an important part of the database system. We don't expect a student's ID number to be empty, and we don't expect the residue capacity of a course in the course selection system to be less than 0. All these can be solve by integrity constrains.

We make integrity constrains as an object in our program. And include it to the local table object. Because integrity constrain is defined when creating a table. Integrity constrain is an interface. Which can check if a update operation can be permitted with giving it a check strategy (it's also an interface) and update operation description.

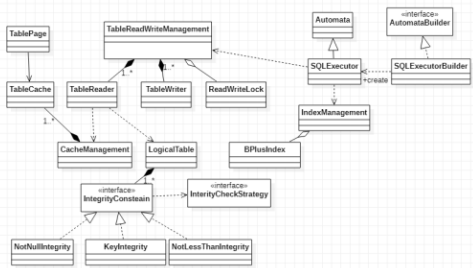

Figure 18. A simple UML class diagram of the several class mentioned.

When an update operation is to be performed, the integrity constraint strategies of the table are called to test integrity constrains. An update operation is performed only if all integrity tests return true.

### 4.6. View

View object is on the top layer, which contain all the data of a query. It implements Serializable interface, so that it can be transferred in the network by using ObjectStream. The SQL Executor will finally return a view object if it's a *select* statement. And the view object will be serialized and return to client.